\documentclass{article}
\usepackage{spconf,amsmath,graphicx,xspace,amssymb,booktabs,multirow,arydshln}

\usepackage[hidelinks]{hyperref}
\usepackage[sort]{cite}
\usepackage{balance}

\newcommand{\tsixty}{$\mathrm{RT}_{60}$\xspace}
\newcommand{\cfifty}{$\mathrm{C_{50}}$\xspace}

\title{Contrastive Representation Learning for Acoustic Parameter Estimation}

\name{Philipp G\"{o}tz$^{1}$, Cagdas Tuna$^{2}$, Andreas Walther$^{2}$, Emanu\"{e}l A. P. Habets$^{1}$\thanks{$^\dag$A joint institution of the Friedrich-Alexander-Universit\"{a}t Erlangen-N\"{u}rnberg (FAU) and Fraunhofer IIS.
\newline Corresponding author: philipp.goetz@audiolabs-erlangen.de.}}
\address{$^1$International Audio Laboratories Erlangen$^\dag$, Germany.
\\ $^2$Fraunhofer Institute for Integrated Circuits IIS, Erlangen, Germany.}

\begin{document}
\ninept
\maketitle
\begin{abstract}
A study is presented in which a contrastive learning approach is used to extract low-dimensional representations of the acoustic environment from single-channel, reverberant speech signals. Convolution of room impulse responses (RIRs) with anechoic source signals is leveraged as a data augmentation technique that offers considerable flexibility in the design of the upstream task. We evaluate the embeddings across three different downstream tasks, which include the regression of acoustic parameters reverberation time \tsixty and clarity index \cfifty, and the classification into small and large rooms. We demonstrate that the learned representations generalize well to unseen data and perform similarly to a fully-supervised baseline.
\end{abstract}


\section{Introduction}
A key challenge in acoustic scene analysis (ASA) is to infer the environmental setting based on an examination of an audio recording \cite{bregman1994auditory}. Within this field of research, there is a wide range of tasks, such as environment classification \cite{eaton2016estimation} or sound event detection \cite{Mesaros2018_TASLP}, whose solutions advance the state-of-the-art in applications such as hearing aid technology \cite{doclo2010acoustic}, audio forensics \cite{radhakrishnan2005audio}, or robot audition \cite{chu2006scene}. In some scenarios, a short audio segment is attributed to a single environmental class, i.e., where the particular sample was recorded \cite{suh2020designing}. In other cases, an audio segment is subject to a time-dependent decomposition into the different contributing sound source types \cite{adavanne2018sound}. Apart from such classification tasks, there are regression tasks that aim to estimate room-acoustic parameters, such as reverberation time \tsixty \cite{gotz2022blind}, or clarity index \cfifty \cite{gamper2020blind}, or geometric properties, like room volume \cite{genovese2019blind}.

A common characteristic in many ASA problems is the highly variant nature of the observed signals. Originating from real-world environments, these signals contain intricate mixtures of various non-stationary sound sources and may vary significantly in duration. Therefore, it may be desirable to represent the signal observed in an unknown acoustic environment by a fixed-size embedding that captures the information of interest while remaining invariant to other information. A similar challenge is encountered in text-independent speaker verification \cite{dehak2010front,snyder2018x}, where a person's identity is represented regardless of the particular utterance and the underlying acoustic conditions. Conversely, the present study aims to extract a low-dimensional representation of the underlying room-acoustic conditions from a single-channel reverberant speech signal. Here, the representation is independent of the speaker's identity and the uttered words. This objective is closely related to acoustic matching \cite{su2020acoustic}, where a reverberant input signal is processed to sound as if recorded in different surroundings. This is achieved by matching the source-independent latent representations of the source and target channels and reconstructing the transformed output signal. While acoustic matching is concerned with generating perceptually plausible signals, our focus is the representation of physically relevant parameters in an embedding from which we can extract information about the acoustic environment.

A central concept of contrastive representation learning \cite{jaiswal2020survey,saeed2021contrastive} is the definition of semantic similarity (positive/negative) between input samples, as it forms the basis of the discrimination task and represents an upstream design criterion that directly determines which information is encoded in the embedding space. This may include position-independent environmental features, such as room volume or reverberation time, or position-dependent quantities, like source-receiver distance or direct-to-reverberant ratio. In the present work, we assume reverberant speech samples that share the same acoustic impulse response but contain different speech signals to form a positive set. In other words, we leverage convolution with anechoic speech recordings as a data augmentation technique and train a softmax classifier to identify samples with the same acoustic transmission path. Following this definition, we encourage the encoder network to learn to discriminate between source and channel information. While this proposed concept of similarity forms the basis of the presented work, the authors note that this definition can be freely extended to other criteria. For this purpose, we test the hypothesis that the consideration of different RIRs from the same room as one class should induce a reduction in downstream performance in position-dependent tasks, e.g., \cfifty estimation, while improving accuracy in position-independent tasks, such as volume classification or \tsixty estimation.
 
\section{Problem formulation}
\label{sec:problem}
A reverberant signal observed in a room can be defined as the convolution of an anechoic source component, e.g., speech from a human speaker or the sound emitted by a machine, with an RIR, which represents a time-domain description of the acoustic transmission path and contains a wealth of information about the environment. In the presence of uncorrelated, additive sensor noise, this definition can be expressed as:
\begin{equation}
    \mathbf{y}=\mathbf{x}\ast\mathbf{h} + \mathbf{n},
\end{equation}
where $(\ast)$ is the convolution operator and $\mathbf{y}$, $\mathbf{x}$, $\mathbf{h}$ and $\mathbf{n}$ denote the observation, the anechoic source signal, the RIR, and uncorrelated, stationary noise, respectively. We train an encoder network to generate a low-dimensional representation $\mathbf{e}\in\mathbb{R}^{1 \times 64}$ from the input $\mathbf{y}$, which encodes only information about the acoustic environment contained in $\mathbf{h}$ and is invariant to the source and noise signals $\mathbf{x}$ and $\mathbf{n}$, respectively. Once upstream convergence is reached, the learned representation serves as input to a downstream network $\text{D}(\mathbf{e})$ that is trained on three separate tasks. The estimation of reverberation time \tsixty in seconds and clarity index \cfifty in decibels, and a binary classification of the room volume.

\section{Proposed method}
\label{sec:method}
The following section covers the contrastive learning approach, the data generation and batch construction, the considered model architecture, and the training procedures.
\subsection{Supervised contrastive loss}
\label{ssec:loss}
One of the advantages of unsupervised learning is the ability to use the abundance of unlabelled data by considering the views of any single sample as semantically different from the views of all other samples. However, in the context of contrastive learning, this approach also induces a central drawback in the possibility of false negatives, which ultimately limits the discrimination performance. To avoid this upper bound in the presented work, we exploit label information and apply a supervised contrastive loss \cite{khosla2020supervised} to the latent representations of the extracted embeddings. Following \cite{chen2020simple}, we use an additional projection network during upstream training that maps $\mathbf{e}$ to a lower-dimensional latent representation $\text{P}(\mathbf{e}) = \mathbf{z}\in\mathbb{R}^{1 \times 16}$ based on which the contrastive loss is computed. The extracted embeddings and their latent representations are $l2$-normalized, such that they represent coordinates on the surfaces of hyper-spheres, where the contrastive loss jointly optimizes the close alignment of similar samples and the approximately uniform distribution of dissimilar samples \cite{wang2020understanding}.

We consider a set $\mathcal{B}$ of $N$ randomly selected RIRs, each augmented with $M$ anechoic speech signals and assigned index $i\in \mathcal{I}\equiv\left\{1,2,\ldots,MN\right\}$. For each sample $\mathbf{y}_i$, the set $\mathcal{A}(i)\equiv \mathcal{I}\setminus\left\{i\right\}$ contains the indices of all other members of $\mathcal{B}$, i.e., all remaining positive and negative samples, with cardinality $|\mathcal{A}|=MN-1$. Furthermore, for each sample $\mathbf{y}_i$, we define the set $\mathcal{P}(i)\equiv\left\{p\in \mathcal{A}(i)\mid \mathbf{y}_p\equiv\mathbf{y}_i\right\}$ to contain the indices of all positive samples relative to $\mathbf{y}_i$. Following \cite{khosla2020supervised}, the supervised contrastive loss is given by:
\begin{equation}
    \mathcal{L_{\mathrm{sup}}} = \sum_{i\in \mathcal{I}} \frac{-1}{|\mathcal{P}(i)|} \sum_{p\in \mathcal{P}(i)} \log \frac{\exp(\langle\mathbf{z}_i,\mathbf{z}_p\rangle/\tau)}{\sum_{a\in \mathcal{A}(i)} \exp(\langle\mathbf{z}_i,\mathbf{z}_a\rangle/\tau) },
    \label{eq:loss}
\end{equation}
where the inner product $\langle\cdot,\cdot\rangle$ represents the cosine similarity between two latent vectors and $\tau$ is a loss temperature hyperparameter that controls the hardness of the estimated class probabilities and reflects the confidence of the classifier.

\subsection{Sampling strategies}
\label{ssec:sampling}
\begin{figure}[t]
    \centering
    \includegraphics[width=\columnwidth]{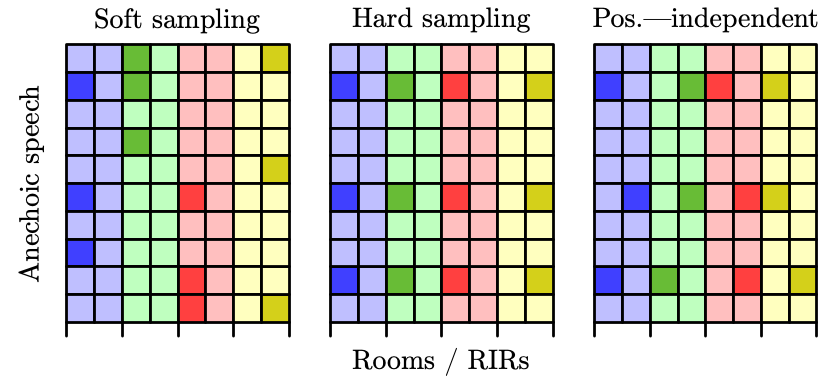}
    \vspace{-2em}
    \caption{Illustration of the multiviewed batch construction from a grid dataset: each square represents an individual reverberant input sample, where the row and column indices represent the anechoic speech components and the RIRs, respectively. In this example, two RIRs are simulated for each of the four generated rooms, indicated by the different colors, while the batch size is $N=4$, and the number of different augmentations per sample is $M=3$.}
    \label{fig:sampling}
\end{figure}

For the training of the upstream encoder, we investigate different batch construction policies that relate to the concept of hard sampling \cite{robinson2020contrastive,kalantidis2020hard}. The informal definition of a hard negative sample pair is one that ``greatly benefits" the model and is highly informative during training. It may be thought of as two observations that appear very similar but belong to different classes and, as such, significantly improve the discrimination ability of the model compared to observations that are easily classified correctly. Translating this definition to our context of acoustic environment embeddings, a hard negative pair consists of two reverberant samples that share the same anechoic speech component but contain different RIRs. As a result, the encoder is presented with data that is semantically different but structurally quite similar. Consequently, the encoder is able to learn  subtle differences between the two instances of the same source signal recorded in different environments. In the presented experiments, we train the upstream encoder network with multiviewed batches that contain RIRs augmented with randomly selected speech signals, referred to as \emph{soft sampling}, and with multiviewed batches, in which the selected RIRs share the same speech signal, referred to as \emph{hard sampling}. Furthermore, we extend the latter approach to include the possibility that for each augmentation a different RIR -- yet from the same room -- is selected. Figure \ref{fig:sampling} shows an illustrative comparison of the sampling strategies.
\begin{figure*}[t]
    \centering
    \includegraphics[width=\textwidth]{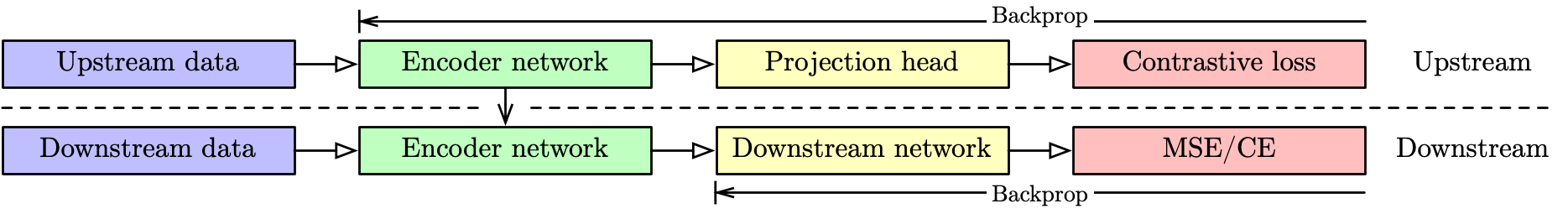}
    \caption{Illustration of the pipeline. During downstream training and evaluation, the weights of the encoder network are kept fixed. Depending on the target, either mean squared error (\tsixty and \cfifty regression) or cross-entropy loss (volume classification) are minimized.}
    \label{fig:pipeline}
\end{figure*}
\subsection{Data generation}
\label{ssec:data}
The present study only considers reverberant speech and assumes stationary acoustic conditions with a single source-microphone transmission path. The scope of this study is limited to rectangular rooms, in which RIRs are generated using an implementation of the image source method provided by \emph{pyroomacoustics} \cite{scheibler2018pyroomacoustics}. Each room has random length, width, and height dimensions in meters, uniformly distributed in the intervals $L\in[3,10]$, $W\in[3,10]$, and $H\in[3,5]$, respectively. The source and the microphone are positioned randomly within the room, while a minimum source--microphone distance and a minimum boundary distance of $50\,\mathrm{cm}$ is ensured. Conveniently, \emph{pyroomacoustics} also contains a database of frequency-dependent absorption coefficients for a wide range of construction materials, such as drywall, carpet or glass. Each of the six boundary surfaces in a room is assigned a randomly selected material from this database, which introduces considerable diversity in the data and adds realism to the synthetic RIRs.
\begin{figure}[b]
    \centering
    \includegraphics[width=\columnwidth]{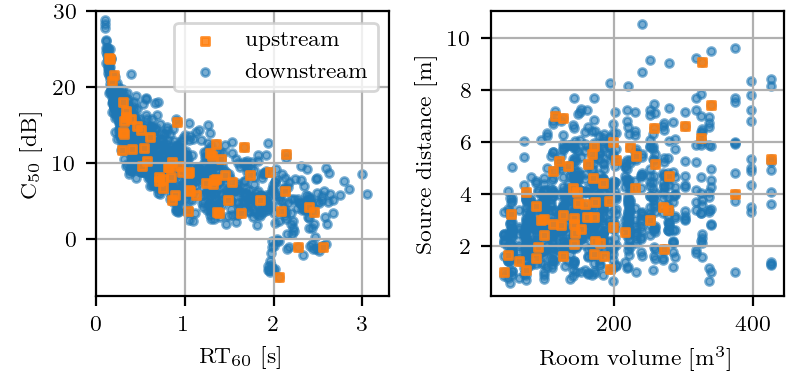}
    \caption{Distribution of acoustic and geometric parameters in the simulated RIRs of the upstream and downstream data sets.}
    \label{fig:gt}
\end{figure}
We make use of the LibriSpeech ASR corpus \cite{panayotov2015librispeech} to construct two fully disjoint datasets. For the upstream training, we simulate RIRs from $64$ different rooms and convolve each instance with $128$, $32$, and $32$ four-second segments of anechoic speech to obtain $8192$, $2048$, and $2048$ reverberant speech samples for the training, validation, and test subsets, respectively. For the downstream dataset, the same number of samples are generated as in the upstream case, with the difference that for each sample, an RIR is randomly selected from a data set of $1000$ RIRs from $100$ different rooms ($10$ RIRs per room). The time-domain signals are transformed via STFT to log-magnitude spectrograms, which are subsequently standardized, i.e., they are zero--mean and normalized to unit variance, which accelerates the training process \cite{lecun2012efficient}. With a sampling rate of $16\,\mathrm{kHz}$, a window size of $32$ samples, and a hop size of $16$ samples, we choose a high temporal resolution that preserves fine acoustic reflection structures in the input signal, which encode geometric cues from the environment. For the supervised downstream training, we compute the \tsixty ground truth based on the energy decay curve \cite{schroeder1965new}. For the binary room volume classification, we define a decision boundary of $160\,\mathrm{m}^{3}$ to ensure that each of the two subsets contains an approximately equal number of samples. The distribution of acoustic and geometric parameters in the two data sets is shown in Figure \ref{fig:gt}.
\subsection{Model architectures}
\label{ssec:model}
We employ a convolutional neural network (CNN) encoder, whose architecture is inspired by prior work in acoustic parameter estimation \cite{gotz2022blind, gamper2018blind}. The encoder network consists of six convolutional blocks, which extract time-frequency features from the input spectrogram. Each convolutional block comprises a convolution layer, a ReLU activation function, and a batch normalization \cite{ioffe2015batch} across the convolution channels. The kernel and stride dimensions of all blocks are listed in Table \ref{tab:conv_layers}. A subsequent dropout regularization prevents overfitting \cite{srivastava2014dropout} before a fully-connected layer projects the compressed input spectrogram to the fixed-size embedding $\mathbf{e}$.
\begin{table}[t]
\centering
{\renewcommand{\arraystretch}{1.25}
\begin{tabular}{r c c c c c c}
\toprule
Layer & $1$ & $2$ & $3$ & $4$ & $5$ & $6$ \\
\hline
Kernel  & $(1, 4)$ & $(1, 4)$ & $(1, 4)$ & $(1, 4)$ & $(2, 4)$ & $(2, 4)$ \\
Stride  & $(1, 2)$ & $(1, 2)$ & $(1, 2)$ & $(1, 2)$ & $(1, 2)$ & $(2, 2)$ \\
\bottomrule
\end{tabular}}
\caption{Parametrization of the six convolutional layers along the dimensions $(\mathrm{frequency}, \mathrm{time})$, no zero-padding was used.}
\label{tab:conv_layers}
\end{table}
For all experiments, an embedding vector of length $64$ was used. As mentioned in Sec.\,\ref{ssec:loss}, we use an additional projection network, which produces the latent representation $\mathbf{z}$ and consists of a single hidden layer with $128$ neurons and a ReLU activation function. The entire upstream model has a total of $157\,\mathrm{k}$ parameters, $2600$ of which are part of the projection network.

The downstream model contains the trained upstream encoder and processes the generated embedding by a sequence of two fully-connected layers, each containing $256$ neurons. After each projection, a non-linear ReLU activation is applied before the hidden representation is mapped to a particular downstream target. In the case of \tsixty, which only assumes values greater than one, a final ReLU activation is added. For the volume classification, a final softmax function yields the predicted probabilities. The downstream section contains $82.7\,\mathrm{k}$ trainable parameters, bringing the total number of the entire model to $237\,\mathrm{k}$ parameters.

\subsection{Training procedure}
\label{ssec:training}
The upstream data set was randomly sampled for each forward pass, we defined $128$ mini-batches to form an epoch and performed a validation over $32$ mini-batches to determine early stopping. If the validation loss did not decrease for four successive epochs, we stopped training and retained the best-performing model. The temperature parameter $\tau$ in (\ref{eq:loss}) controls the hardness of the estimated class distribution and, thereby, the distinctness of the produced embeddings. We trained the upstream encoder at three different temperature coefficients $\tau\in\left\{0.01,\,0.1,\,1.0\right\}$ to investigate the dependence of the learned representations on this hyperparameter. We used a batch size of $N=24$ with $M=3$ augmentations per sample and updated the model weights with a learning rate of $10^{-3}$ using the Adam optimizer \cite{kingma2014adam}. The downstream model was trained with every upstream encoder on the three different tasks separately, with a learning rate of $10^{-3}$ and a batch size of $16$. The dropout operation between the convolutional block and the fully-connected layer in the upstream encoder was disabled during downstream training, except for the supervised baseline.

\section{Evaluation}
\label{sec:eval}
The different sampling strategies and loss temperatures were evaluated based on common performance metrics, which include root-mean-square error (RMSE), Pearson's correlation coefficient (CORR) and estimation bias for the \tsixty and \cfifty regression, and accuracy (ACC), precision (PR) and recall (RE) for the volume classification.

\begin{table*}[htb!]
    \centering
    {\renewcommand{\arraystretch}{1.0}
    \begin{tabular}{l l @{\hspace{0.75cm}} c c c c c c c c c c c}
    \toprule
         & & \multicolumn{3}{c}{\textsc{Soft sampling}} && \multicolumn{3}{c}{\textsc{Hard sampling}} && \multicolumn{3}{c}{\textsc{Pos.--independent}}  \\
         \cmidrule{3-5} \cmidrule{7-9} \cmidrule{11-13}
         & & RMSE & CORR & BIAS && RMSE & CORR & BIAS && RMSE & CORR & BIAS \\
         \midrule

         \multirow{3}{*}{\tsixty $[\mathrm{s}]$} & Supervised & $0.2228$ & $0.9037$ & $0.0304$ && $0.2228$ & $0.9037$ & $0.0304$ && $0.2228$ & $0.9037$ & $0.0304$ \\
         
         & $\tau=0.01$ & $0.2699$ & $0.8496$ & $0.0344$ && $0.2369$ & $0.8870$ & $0.0322$ && $0.2146$ & $0.9089$ & $0.0296$ \\
         
         & $\tau=0.1$ & $0.2345$ & $0.8878$ & $0.0027$ && $0.2426$ & $0.8787$ & $0.0069$ && $0.2217$ & $0.9048$ & $0.0028$ \\
         
         & $\tau=1.0$ & $0.2500$ & $0.8768$ & $0.0450$ && $0.2770$ & $0.8399$ & $\hspace{-7pt}-0.0043$ && $0.2284$ & $0.8937$ & $0.0220$ \\
         \midrule
         \multirow{3}{*}{\cfifty $[\mathrm{dB}]$} & Supervised & $1.9756$ & $0.9241$ & $1.0281$ && $1.9756$ & $0.9241$ & $1.0281$ && $1.9756$ & $0.9241$ & $1.0281$ \\
         
         & $\tau=0.01$ & $1.7561$ & $0.9208$ & $0.3187$ && $2.0057$ & $0.9227$ & $0.9820$ && $2.2386$ & $0.8908$ & $1.0001$ \\
         
         & $\tau=0.1$ & $1.8903$ & $0.9223$ & $0.4303$ && $2.0411$ & $0.9142$ & $0.8401$ && $2.0813$ & $0.9063$ & $0.8936$ \\
         
         & $\tau=1.0$ & $1.7887$ & $0.9181$ & $0.3901$ && $2.0101$ & $0.9032$ & $0.7059$ && $2.3873$ & $0.8597$ & $0.7434$ \\[1em]
         & & ACC [$\%$] & PR [$\%$] & RE [$\%$] && ACC [$\%$] & PR [$\%$] & RE [$\%$] && ACC [$\%$] & PR [$\%$] & RE [$\%$] \\
         
         \midrule
         \multirow{3}{*}{Volume $[\mathrm{m}^3]$} & Supervised & $68.95$ & $70.66$ & $70.16$ && $68.95$ & $70.66$ & $70.16$ && $68.95$ & $70.66$ & $70.16$ \\
         
         & $\tau=0.01$ & $69.53$ & $71.46$ & $71.03$ && $70.70$ & $72.52$ & $71.99$ && $74.22$ & $73.20$ & $73.28$ \\
         
         & $\tau=0.1$ & $71.29$ & $72.41$ & $71.65$ && $67.97$ & $70.14$ & $69.92$ && $73.05$ & $75.86$ & $76.02$ \\
         
         & $\tau=1.0$ & $71.09$ & $73.76$ & $73.83$ && $67.58$ & $69.43$ & $69.05$ && $68.36$ & $70.24$ & $69.84$ \\
         \bottomrule
    \end{tabular}}
    \caption{Downstream evaluation metrics. The results for the supervised baseline are repeated in all three sampling columns for convenience.}
    \label{tab:results}
\end{table*}

The results listed in Table \ref{tab:results} show that the representations learned from the contrastive training are highly suitable as input in solving the three downstream tasks. While the results for the different sampling strategies and loss temperatures are generally similar, in each of the considered tasks, there is a remarkable trend that the models trained with a contrastive loss, without additional fine-tuning, perform at least on par with the fully supervised baseline. In a short experiment, intended as a proof of concept, we trained a downstream model that contained an untrained upstream encoder to estimate \tsixty, and achieved rather large errors of $0.5056\,\mathrm{s}$ and low correlation coefficients of $0.1993$.

We confirm our hypothesis that the position-independent sampling degrades the performance for the position-dependent \cfifty while it improves the performance for the \tsixty and the room volume classification. This effect is reflected by an increase in RMSE of around $14\%$ for \cfifty and an accuracy improvement of the volume classification from the soft to the position-independent sampling. A possible explanation for this rather moderate change in performance can be found in the ground truth distribution (cf. Fig.\,\ref{fig:gt}), which shows the correlation between \tsixty and \cfifty. If the learned representations are, by definition, position-independent, the downstream model may infer \cfifty from a well-estimated \tsixty.

\begin{figure}[!t]
    \centering
    \includegraphics[width=\columnwidth]{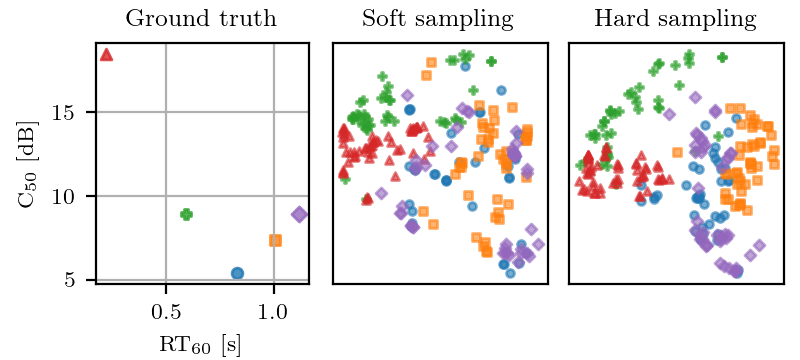}
    \caption{UMAP plot of the embeddings generated from speech samples in five different rooms. The plot on the left shows the distribution of \tsixty and \cfifty, and the center and right plots show the UMAP projections for the soft and hard sampling strategies. There is a noticeable difference in the degree of the clustering of samples that belong to the same room.}
    \label{fig:emb}
\end{figure}

The temperature coefficients used during upstream training do not systematically affect the downstream performance. The results indicate that an optimal choice may depend on the specific task. A further investigation with a finer sampling of the temperature scale may offer additional insights. The effect of hard sampling, as outlined in Sec.\,\ref{ssec:sampling}, can be seen in Fig.\,\ref{fig:emb}, where we use uniform manifold approximation \cite{mcinnes2018umap} to qualitatively compare the embeddings of reverberant speech samples from five different rooms. We confirm our hypothesis that sharing the maximum number of speech samples per batch increases the degree of invariance to the source component and results in representations with an increased focus on the channel information. Thus, in the case of hard sampling, the discrimination task during upstream training is simplified, as the model does not have to distinguish between different speech samples in addition to the different RIRs. 

The authors want to emphasize that the results shown in this work represent an exploratory study of our proposed approach to learning room-acoustic representations from reverberant speech signals. We acknowledge that the basic concept of attributing similarity to different observations may be extended to other criteria, depending on the requirements of the representations, which are dictated by the specific downstream task. We hope to inspire further work that adapts our approach to meet different requirements.

\section{Conclusion}
\label{sec:conc}
We presented an approach to extract low-dimensional representations from reverberant speech signals in a contrastive learning framework. Different batch construction strategies were proposed and their effectiveness across three well-known downstream tasks of acoustic parameter estimation and environment classification were demonstrated. By performing at least on par with a fully-supervised baseline, we conclude that the proposed method yields a rich representation of the acoustic environment while being invariant to characteristics relating to the source signal.

\balance
\bibliographystyle{IEEEbib}
\bibliography{short-strings,refs}

\end{document}